\documentclass[preprint]{acm_proc_article-sp}
\usepackage[utf8]{inputenc}
\usepackage{booktabs}
\usepackage{textcomp}
\usepackage{diagbox}
\usepackage{multirow}

\newcommand{\gp}{$G_{R,\theta}$\-}

\newenvironment{mycaption}{\sffamily\small}{}

\begin{document}

\title{Comparing the sensitivity of social networks, web graphs, and random graphs with respect to vertex removal}

\numberofauthors{2}\author{
\alignauthor
Christoph Martin\\
       \affaddr{Leuphana University,}
       \affaddr{Scharnhorststra{\ss}e 1 }\\
       \affaddr{21335 L\"uneburg, Germany}\\
       \email{cmartin@leuphana.de}
\alignauthor
Peter Niemeyer\\
       \affaddr{Leuphana University,}
       \affaddr{Scharnhorststra{\ss}e 1 }\\
       \affaddr{21335 L\"uneburg, Germany}\\
       \email{niemeyer@uni.leuphana.de}
}

\maketitle
\begin{abstract}

The sensitivity of networks regarding the removal of vertices has been studied extensively within the last 15 years.  A common approach to 
measure this sensitivity is (i) removing successively vertices by following a specific removal strategy and (ii) comparing the original and the 
modified network using a specific comparison method.

In this paper we apply a wide range of removal strategies and comparison methods in order to study the sensitivity of medium-sized networks 
from real world and randomly generated networks.

In the first part of our study we observe that social networks and web graphs differ in sensitivity. When removing vertices, social networks
are robust, web graphs are not. This effect is conclusive with the work of Boldi et al. who analyzed very large networks.

For similarly generated random graphs we find that the sensitivity highly depends on the comparison method. The choice of the removal strategy 
has surprisingly marginal impact on the sensitivity as long as we consider removal strategies implied by common centrality measures. However, 
it has a strong effect when removing the vertices in random order.

\end{abstract}

\category{}{}{}

\terms{}

\newpage

\section{Introduction}

Networks are part of our everyday life -- we are in contact with social networks and unconsciously interact with web graphs every day. Although these types of networks represent completely different constructs, they share various structural properties (e.g. heavy-tailed degree distributions, short average distances).
Recently, Boldi et al. \cite{Boldi2013a} observed for very large networks that social networks and web graphs behave inherently different 
under controlled vertex removal. 
While social networks appear to be robust, web graphs are very sensitive to certain modifications.

To measure the sensitivity of a graph, we (i) successively remove vertices following a specific removal strategy and (ii) compare the original and the 
modified networks using a specific comparison method based on either the shortest path distribution or a centrality measure.

Measuring the sensitivity by comparing modified graphs to their respective source graph is a common concept: In the field of social network 
analysis, sampling errors are simulated to judge the robustness of centrality measures \cite{Bolland1988a, Costenbader2003a, Borgatti2006a, 
Wang2012a}. In web science, networks are modified in a controlled way to evaluate their 
vulnerability against attacks \cite{Albert2002a, Holme2002a, Boldi2013a, Iyer2013a}.

In this study, we analyze the sensitivity of graphs with respect to vertex removal induced by removal strategies.
The removal strategy defines the order by that vertices are removed from the network. In this paper, we discuss removal strategies induced by centrality measures (i.e.: first remove vertices with high centrality values, e.g. degree centrality), as well as a removal strategy based on a community detection algorithm (label propagation).
The comparison method defines how to compare modified and unmodified networks. In addition to comparison methods based on the neighborhood function (as applied in \cite{Boldi2013a, Cabral2014a}), we consider comparison methods based on centrality measures (here we measure the rank correlation between centrality measures of the modified and the unmodified network).

The main contribution of this paper is twofold: First, we analyze the sensitivity of 
medium-sized real-world networks (in contrast to previous 
studies on small and very large networks) and confirm previous results. Second, we 
observe that randomly generated networks  (Erd\H{o}s-R\'enyi model, Bara\-basi-Albert model, Watts-Strogatz model, and configuration model) behave differently depending on whether the vertices are removed in random order or by a removal strategy implied by centrality measures.

\section{Related work}
Modifying graphs and comparing the outcome with the respective source graph is a common approach to tackle a variety of questions. In the field of social 
network analysis, networks are modified in order to simulate measurement errors and to examine the robustness of centrality measures. An empirical network was altered by Bolland \cite{Bolland1988a} and Pearson correlation was used to measure the robustness. Random samples have been taken 
from empirical networks  by Costenbader and Valente \cite{Costenbader2003a} to investigate the stability of various centrality measures. Four error types have been applied to  
Erd\H{o}s-R\'enyi graphs by  Borgatti et al. \cite{Borgatti2006a} to measure centrality robustness regarding different types of accuracy measures. Six types of measurement 
errors have been applied to real-world and generated networks by Wang et al. \cite{Wang2012a} in order to examine the robustness of node-level network measures by 
means of Spearman's rho. 

Albert and Barabási \cite{Albert2002a} examined the error and attack tolerance of random graphs with respect to node 
removal. Based on centrality measures, Holme et al. \cite{Holme2002a} removed nodes and edges from real-world networks and random graphs to investigate the attack 
vulnerability regarding the average inverse of the geodesic length and the size of the largest connected subgraph. The behavior of very large social networks 
and web graphs with respect to vertex removal based on various removal strategies and comparison methods based on the neighborhood function has 
been studied by Boldi et al. \cite{Boldi2013a}. Cabral et al. \cite{Cabral2014a} measured the impact of random errors to real-world networks and generated graphs by means of 
stochastic quantifiers.

In our study, we combine several techniques: based on various removal strategies, we modify medium-sized real-world networks and random graphs 
and evaluate the sensitivity of those 
networks on the basis of  the shortest path distribution, stochastic quantifiers, and centrality measures. 

\section{Concepts}
A graph $G(V,E)$ is represented by a set of nodes $V$ with $|V| = n$ and a set of edges $E$ with $|E| = m$. All used graphs in this work are 
unweighted and either directed or undirected. 
In this work, the terms graph and network are used interchangeably.
The neighborhood function $N$ of a graph $G$ at $t$ is the number of pairs of nodes within distance~$t$:
\begin{equation}
N_G(t) = | \{(u,v) : u \in V, v \in V, dist(u,v) \leq t\}|,
\end{equation}
with $dist(u,v)$ as the geodesic distance between $u$ and $v$ \cite{Palmer2002a}.\footnote{In the case of an undirected graph: 
\begin{math}
N_G(t) = | \{\{u,v\}: u \in V, v \in V, dist(u,v) \leq t\}|
\end{math}.}

The neighborhood function can be approximated. Hence, we are capable of calculating $N$ for graphs where the exact calculation of $N$ has
infeasible running time. Multiple approximation algorithms exist \cite{Palmer2002a, Boldi2011a}, we use HyperANF \cite{Boldi2011a}.

A multitude of measures such as the number of reachable pairs \cite{Boldi2013a} and the average path length is derived from $N$.
Besides, we are specifically interested in the harmonic diameter \cite{Marchiori2000a} which is defined as follows:

\begin{align}
D_{harm}(G) &= \frac{n ( n-1)}{\sum_{u \neq v} (dist(u,v))^{-1}} \\
&= \frac{n(n-1)}{\sum_{t > 0} \frac{1}{t} (N_{G}(t) - N_{G}(t-1))} 
\end{align}

Moreover, we derive the number of shortest paths at distance $t$ from $N$:
\begin{equation}
SP_G(t) =
\begin{cases}
N_G(t) &\mbox{if } t = 0 \\
N_G(t) - N_G(t-1) &\mbox{if } t > 0 \\
\end{cases}
\end{equation}
Thus, we respresent the probability mass function of the shortest path distribution:
\begin{equation}
H_{SP_G}(t) = \frac{SP_G(t)}{\sum_{t} SP_G(t) }
\end{equation}

\subsection{Graph modification and removal strategies}
\begin{figure}[h]
\caption{Procedure to measure the sensitivity}
\label{fig:vorgehen}
\includegraphics[width=\linewidth]{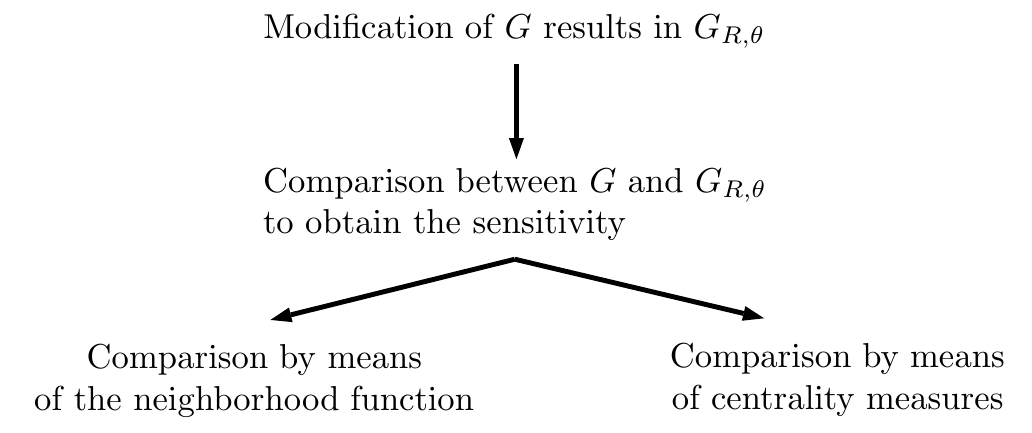}
\end{figure}

Our basic approach is illustrated in Figure~\ref{fig:vorgehen}.  To obtain the modified graph \gp, we
apply a removal strategy $R$ at a certain modification level $\theta$ to a source graph $G$. 

Following Boldi et al. \cite{Boldi2013a}, a removal strategy $R$ specifies the order in which the nodes are removed. We use removal strategies based on centrality
measures and label propagation. In the first case, the
nodes are ordered (descending) by their corresponding centrality value, whereby the centrality measure either is
betweenness centrality $(bc)$, closeness centrality $(cc)$ \cite{Freeman1978a}, degree centrality $(dc)$, eigenvector centrality $(ec)$ \cite{Bonacich1987a}, or PageRank
$(pr)$ \cite{Brin1998a}. For directed graphs, we also use the in and out versions of $cc$ and $dc$ as removal strategy.

The modification level $\theta$ indicates the fraction of edges we remove from the source graph. More
precisely, nodes are removed from the source graph $G$ based on the chosen removal strategy $R$ until $\theta m$ edges are removed. The outcome of this procedure is the modified graph \gp.

The label propagation $(lp)$ removal strategy is based on the label propagation
community detection algorithm \cite{Raghavan2007a, Boldi2013a}. For each cluster, in decreasing size of order the
node with the highest number of neighbors in other clusters is removed. If the first node has been removed in every
cluster and $\theta m$ edges have not been removed yet, the second, third etc. node with the highest number of 
neighbors in other clusters is removed. When centrality measures function as comparison methods, we also use a random removal strategy where every vertex is removed with equal probability.

\subsection{Comparing $G$ and \gp}
After creating the modified graph, we compare $G$ and \gp. Pursuing two different approaches,  based on the neighborhood function and on 
centrality measures, we measure the structural change and thus the sensitivity.

\textbf{Comparison based on the neighborhood function}\\
When comparing $N_G$ and $N_{G_{R,\theta}}$, we make use of the relative harmonic diameter change $\delta$:

\begin{equation}
\delta(G, G_{R, \theta}) = \frac{ D_{harm}(G_{R,\theta}) }{ D_{harm}(G) } -1
\end{equation}

This measure combines information about the path length and the connectivity. It has shown the best performance as comparison method for 
neighborhood functions in \cite{Boldi2013a}.

The use of stochastic quantifiers is another way of comparing neighborhood functions.
In this work, we use the same
quantifiers as Cabral et al. \cite{Cabral2014a}, specifically the Kullback-Leibler divergence ($kl$), the Jensen-Shannon distance 
($jsd$) as well as the Hellinger distance ($hd$):
\begin{equation}
\label{eq:kl}
kl(G, G_{R, \theta}) = kl(H_{SP_G}, H_{SP_{G_{R, \theta}} })
\end{equation}
Analogously to Equation~\ref{eq:kl} we define $jsd(G, G_{R, \theta})$ and \\ $hd(G, G_{R, \theta})$.

\textbf{Comparison based on centrality measures}\\
The second approach to compare the network structure of $G$ and \gp\ uses a centrality measure $cm \in (bc, cc, dc, 
ec, pr)$. Since \gp\ is an induced subgraph of $G$, the centrality values for every node $u$ in \gp\ are calculated for both graphs. The results are stored in the vector $M$ for $G$ and $M_{R,
\theta}$ for \gp. We measure the sensitivity by computing Spearman's rank correlation coefficient $\rho$:
\begin{equation}
\rho(G, G_{R,\theta}) = \rho(M, M_{R,\theta})
\end{equation}

This approach is common in the field of robustness of network measures (cf. Wang et al. \cite{Wang2012a}).

\subsection{Random graph models}
\label{sec:randomGraphs}
In section~\ref{sec:generated}, we apply our procedure to graphs generated by the following 
random graph models:

According to the \textit{Erd\H{o}s-R\'enyi model} ($ER(n,p)$), a graph consists of $n$ nodes.
The existence of an edge between two nodes is specified by  the probability $p$. Subsequently, the degree distribution follows a binomial distribution \cite{Erdos1959a}.

As introduced in \cite{Albert2002a}, the \textit{Barabasi-Albert model} ($BA(n,l)$)  is based on the assumption that a network grows over time. The initial network
consists of a single node. In each time step a new node is added and connected to $l$ other nodes
chosen from the existing nodes with a probability proportional to their degree.
New nodes are added until the graph consists of $n$ nodes. The networks 
generated by the $BA$ model follow a power-law degree distribution.

Following the \textit{Watts-Strogatz small-world model} ($WS(n,k,p_{rew})$) \cite{Watts1998a}, the initial graph is a ring with $n$ nodes that are 
connected to $k$ predecessors and successors. Afterwards, each edge is randomly rewired with probability $p_{rew}$,
self-loops and multiple edges that may arise are deleted.
 Graphs generated by the $WS$ model exhibit small-world properties, i.e. high transitivity and relatively small average path 
length.

To generate graphs based on a given degree sequence of a graph $G$, the \textit{configuration model} ($CF(G)$) is used \cite[p. 434 ff.]{Newman2010a}.
Initially, every node $v_i$ has $k_i$ stubs ($k_i$ is the degree of the $ith$ node).  Each step, two random stubs are chosen and connected with each other until all stubs are
connected. Self-loops and multiple edges that may arise are deleted.

For graph generation and modification we use the igraph library \cite{Csardi2006a}.

\section{Study of real-world networks}
In this section, we apply our previously described approach to real-world networks. After characterizing the experimental design and the specific 
networks we use, we discuss our results.

\subsection{Experimental design}
Applying our procedure (Figure~\ref{fig:vorgehen}) to the networks described in chapter~\ref{sec:realworlddata} we use $\theta \in \{0.05, 0.1, 0.15, 0.2, 0.25, 0.3\}$. The neighborhood function is calculated exactly for the Hamsterster and Google network. HyperANF is used to approximate 
the neighborhood function in the remaining cases. In case of approximation, we make at least ten runs with 1024 registers per counter to 
ensure relative standard deviations at a maximum of 1.45\% \cite{Boldi2013a}.

For the comparison based on centrality measures, we use $dc$, $ec$ and $pr$. Due to their time complexity, $cc$ and $bc$ are not 
considered as comparison methods. 
The in and out versions of $cc$ and $dc$ are not considered as comparison methods as well since by definition these measures are only available for directed graphs.

\subsection{Data}
\label{sec:realworlddata}
In this section, we use six real-world networks of various sizes represented by three social networks and three web graphs:\footnote{
Multiple edges and self-loops are removed from all graphs.
}
\begin{description}
\item[Hamsterster]
(2.426 nodes, 16.631 edges, undirected): Hamsterster.com was a virtual hamster and gerbil community. The users are connected by edges if 
they share a friendship or family relationship. (available through \cite{Kunegis2013a})
\item[Brightkite]
(58.228 nodes, 214.078 edges, undirected): Bright\-kite.com was a location-based social network. The users are connected when a friendship exists in both directions. This network was created by \cite{Cho2011a}.
\item[Slashdot]
(82.168 nodes, 948.464 edges, directed) Slashdot.com is a technology related news website where users can tag each other as friends or 
foes. This snapshot (february 2009) of the network contains friend/foe links between the users. This network was created by 
\cite{Leskovec2009a}.

\item[Google] (15.763 nodes, 171.206 edges, directed) A web graph based on google.com also used by \cite{Palla2007a}.

\item[Stanford] (281.903 nodes, 2.312.497 edges, directed) A web graph based on the website of the Stanford University (stanford.edu). This 
network is also used in \cite{Leskovec2009a}.

\item[NotreDame] (325.729 nodes, 1.497.134 edges, directed) A web graph based on the website of the University of Notre Dame (nd.edu). 
This network was created by \cite{Albert1999a}.
\end{description}

\subsection{Results}
Especially interested in the behavior
of web graphs compared to social networks, we analyze if web graphs and social networks behave differently under
controlled modification. Our results regarding $\delta$ and $hd$ are illustrated in Table~\ref{tab:resultsEmpirical}.
Considering the comparison based on the neighborhood function, our experiments show that the 
observed social networks are slightly affected by the modification whereas all web graphs are substantially disturbed.
In contrast, the comparison by means of centrality measures does not provide a clear-cut distinction between web graphs and 
social networks. Following, we describe the results in a more detailed manner.

	\begin{table*}
  \caption{Sensitivity of real-world networks with regard to systematic vertex removal (comparison based on the neighborhood function)}
     
\begin{tabular*}{1\textwidth}{@{\extracolsep{\fill} }  llrrrrrrrrrrrr}
    \toprule
    & \multirow{ 2}{*}{ \diagbox{$R$}{$\theta$}} &   \multicolumn{2}{c}{0.05}       & \multicolumn{2}{c}{0.10}       & \multicolumn{2}{c}{0.15}       & \multicolumn{2}{c}{0.20 }& \multicolumn{2}{c}{0.25} & \multicolumn{2}{c}{0.30}\\
    $G$      &     & $hd$  & $\delta$ & $hd$  & $\delta$  & $hd$  & $\delta$  & $hd$  & $\delta$  & $hd$  & $\delta$  & $hd$  & $\delta$ \\
    \toprule
    
    Brightkite & $bc$ & 0.09  & 0.10  & 0.14  & 0.16  & 0.19  & 0.23  & 0.24  & 0.32  & 0.30  & 0.39  & \textbf{0.36}  & \textbf{0.52} \\
          & $cc$ & 0.08  & 0.07  & 0.12  & 0.14  & 0.19  & 0.20  & 0.24  & 0.30  & 0.29  & 0.38  & 0.35  & 0.49 \\
          & $dc$ & 0.07  & 0.08  & 0.12  & 0.12  & 0.16  & 0.17  & 0.19  & 0.23  & 0.23  & 0.30  & 0.29  & 0.39 \\
          & $ec$ & 0.00  & 0.00  & 0.04  & 0.07  & 0.07  & 0.11  & 0.12  & 0.19  & 0.18  & 0.31  & 0.22  & 0.41 \\
          & $lp$ & 0.02  & 0.10  & 0.02  & 0.13  & 0.03  & 0.09  & 0.06  & 0.12  & 0.08  & 0.13  & 0.11  & 0.17 \\
          & $pr$ & 0.08  & 0.07  & 0.13  & 0.16  & 0.18  & 0.23  & 0.22  & 0.29  & 0.27  & 0.38  & 0.32  & 0.47 \\
             \midrule                       
    Hamsterster & $bc$ & 0.05  & 0.04  & 0.08  & 0.08  & 0.10  & 0.15  & 0.12  & 0.20  & 0.16  & 0.28  &\textbf{0.19}  & \textbf{0.35} \\
          & $cc$ & 0.03  & 0.03  & 0.05  & 0.06  & 0.09  & 0.09  & 0.12  & 0.12  & 0.14  & 0.17  & 0.17  & 0.22 \\
          & $dc$ & 0.04  & 0.04  & 0.06  & 0.06  & 0.08  & 0.10  & 0.10  & 0.13  & 0.13  & 0.18  & 0.17  & 0.25 \\
          & $ec$ & 0.03  & 0.03  & 0.04  & 0.04  & 0.06  & 0.07  & 0.08  & 0.09  & 0.10  & 0.14  & 0.13  & 0.19 \\
          & $lp$ & 0.02  & -0.01 & 0.04  & -0.17 & 0.04  & -0.25 & 0.04  & -0.31 & 0.08  & -0.32 & 0.06  & -0.32 \\
          & $pr$ & 0.04  & 0.04  & 0.06  & 0.06  & 0.08  & 0.11  & 0.11  & 0.15  & 0.15  & 0.21  & 0.18  & 0.27 \\
          \midrule 
    Slashdot & $bc$ & 0.05  & 0.06  & 0.08  & 0.11  & 0.10  & 0.16  & 0.13  & 0.22  & 0.15  & 0.27  & 0.19  & \textbf{0.32} \\
          & $cc$ & 0.04  & 0.04  & 0.07  & 0.10  & 0.10  & 0.13  & 0.11  & 0.21  & 0.14  & 0.22  & 0.17  & 0.26 \\
          & $cc_{in}$ & 0.03  & 0.05  & 0.06  & 0.08  & 0.08  & 0.08  & 0.11  & 0.10  & 0.12  & 0.16  & 0.16  & 0.17 \\
          & $cc_{out}$ & 0.05  & 0.06  & 0.07  & 0.10  & 0.09  & 0.14  & 0.12  & 0.17  & 0.13  & 0.22  & 0.17  & 0.27 \\
          & $dc$ & 0.04  & 0.08  & 0.07  & 0.13  & 0.10  & 0.15  & 0.12  & 0.20  & 0.15  & 0.22  & 0.18  & 0.28 \\
          & $dc_{in}$ & 0.04  & 0.09  & 0.07  & 0.12  & 0.09  & 0.16  & 0.12  & 0.20  & 0.15  & 0.23  & 0.18  & 0.28 \\
          & $dc_{out}$ & 0.05  & 0.09  & 0.07  & 0.13  & 0.10  & 0.15  & 0.12  & 0.20  & 0.15  & 0.25  & 0.17  & 0.29 \\
          & $ec$ & 0.03  & 0.04  & 0.05  & 0.07  & 0.08  & 0.12  & 0.10  & 0.17  & 0.13  & 0.20  & 0.16  & 0.26 \\
          & $lp$ & 0.02  & 0.04  & 0.03  & 0.10  & 0.05  & 0.13  & 0.10  & 0.05  & 0.16  & -0.01 & \textbf{0.20}  & -0.06 \\
          & $pr$ & 0.05  & 0.09  & 0.07  & 0.12  & 0.10  & 0.16  & 0.12  & 0.20  & 0.14  & 0.27  & 0.18  & 0.30 \\
          \midrule 
    Google & $bc$ & 0.11  & 0.18  & 0.34  & 0.68  & 0.35  & 0.77  & 0.35  & 0.90  & 0.32  & 2.25  & 0.29  & 6.02 \\
          & $cc$ & 0.09  & 0.07  & 0.10  & 0.07  & 0.13  & 0.11  & 0.15  & 0.19  & 0.19  & 0.33  & 0.22  & 0.37 \\
          & $cc_{in}$ & 0.09  & 0.07  & 0.10  & 0.07  & 0.10  & 0.08  & 0.15  & 0.23  & 0.29  & 0.35  & \textbf{0.50}  & 0.66 \\
          & $cc_{out}$ & 0.01  & 0.01  & 0.32  & 2.82  & 0.31  & 2.91  & 0.35  & 3.28  & 0.21  & 3.83  & 0.21  & 3.83 \\
          & $dc$ & 0.09  & 0.07  & 0.10  & 0.07  & 0.10  & 0.08  & 0.14  & 0.13  & 0.15  & 0.37  & 0.22  & 0.70 \\
          & $dc_{in}$ & 0.09  & 0.07  & 0.10  & 0.07  & 0.10  & 0.08  & 0.14  & 0.13  & 0.13  & 0.19  & 0.25  & 0.42 \\
          & $dc_{out}$ & 0.16  & 0.49  & 0.32  & 6.45  & 0.31  & 6.61  & 0.29  & 7.71  & 0.29  & 7.77  & 0.28  & \textbf{8.51} \\
          & $ec$ & 0.00  & 0.01  & 0.01  & 0.01  & 0.10  & 0.08  & 0.09  & 0.09  & 0.10  & 0.10  & 0.13  & 0.18 \\
          & $lp$ & 0.09  & 0.07  & 0.14  & 0.41  & 0.14  & 0.41  & 0.18  & 0.71  & 0.49  & 1.20  & \textbf{0.50}  & 1.24 \\
          & $pr$ & 0.09  & 0.07  & 0.10  & 0.07  & 0.10  & 0.08  & 0.16  & 0.14  & 0.22  & 0.39  & 0.40  & 0.93 \\
          \midrule 
    NotreDame & $bc$ & 0.39  & 4.31  & 0.50  & 11.46 & 0.62  & 34.48 & 0.53  & 138.50 & 0.50  & 211.12 & 0.49  & \textbf{241.81} \\
          & $cc$ & 0.57  & 4.74  & 0.53  & 16.47 & 0.70  & 36.00 & 0.75  & 37.75 & 0.75  & 41.74 & 0.76 & 35.60 \\
          & $cc_{in}$ & 0.02  & -0.01 & 0.02  & -0.01 & 0.02  & -0.03 & 0.02  & -0.02 & 0.02  & -0.05 & 0.02  & -0.03 \\
          & $cc_{out}$ & 0.50  & 3.16  & 0.56  & 15.69 & 0.69  & 23.88 & 0.77  & 23.51 & 0.78  & 22.76 & \textbf{0.80}  & 22.33 \\
          & $dc$ & 0.20  & 0.61  & 0.23  & 1.23  & 0.26  & 1.51  & 0.27  & 2.09  & 0.27  & 2.12  & 0.27  & 2.06 \\
          & $dc_{in}$ & 0.19  & 0.47  & 0.19  & 0.90  & 0.21  & 1.56  & 0.21  & 1.53  & 0.21  & 1.54  & 0.23  & 1.90 \\
          & $dc_{out}$ & 0.03  & 0.25  & 0.07  & 0.68  & 0.07  & 0.66  & 0.06  & 0.70  & 0.09  & 0.82  & 0.08  & 1.42 \\
          & $ec$ & 0.01  & 0.01  & 0.01  & -0.01 & 0.02  & 0.00  & 0.02  & 0.00  & 0.03  & -0.01 & 0.02  & -0.01 \\
          & $lp$ & 0.22  & 0.88  & 0.26  & 1.76  & 0.30  & 4.00  & 0.31  & 11.84 & 0.33  & 24.09 & 0.36  & 36.89 \\
          & $pr$ & 0.19  & 0.60  & 0.17  & 0.91  & 0.20  & 1.87  & 0.35  & 3.76  & 0.39  & 6.73  & 0.40  & 14.88 \\
          \midrule 
    Stanford & $bc$ & 0.13  & 0.76  & 0.27  & 1.78  & 0.31  & 3.34  & 0.37  & 4.93  & 0.36  & 9.05  & \textbf{0.47}  & \textbf{14.18} \\
          & $cc$ & 0.14  & 0.14  & 0.20  & 0.34  & 0.20  & 0.34  & 0.29  & 0.65  & 0.32  & 1.27  & 0.32  & 1.54 \\
          & $cc_{in}$ & 0.03  & -0.06 & 0.05  & -0.12 & 0.06  & -0.17 & 0.10  & -0.10 & 0.22  & 0.06  & 0.34  & 0.56 \\
          & $cc_{out}$ & 0.03  & -0.03 & 0.08  & 0.05  & 0.28  & 0.48  & 0.32  & 0.52  & 0.33  & 0.54  & 0.38  & 0.65 \\
          & $dc$ & 0.08  & 0.18  & 0.10  & 0.32  & 0.20  & 0.59  & 0.22  & 0.85  & 0.24  & 0.99  & 0.27  & 1.14 \\
          & $dc_{in}$ & 0.08  & 0.18  & 0.10  & 0.33  & 0.20  & 0.62  & 0.21  & 0.83  & 0.24  & 0.99  & 0.27  & 1.10 \\
          & $dc_{out}$ & 0.06  & 0.13  & 0.13  & 0.50  & 0.16  & 0.75  & 0.19  & 1.03  & 0.22  & 1.33  & 0.24  & 1.82 \\
          & $ec$ & 0.07  & 0.10  & 0.20  & 0.36  & 0.27  & 0.58  & 0.39  & 1.12  & 0.41  & 3.33  & 0.39  & 4.65 \\
          & $lp$ & 0.15  & 0.36  & 0.20  & 1.20  & 0.24  & 2.97  & 0.22  & 4.86  & 0.26  & 6.54  & 0.14  & 8.49 \\
          & $pr$ & 0.14  & 0.36  & 0.18  & 0.52  & 0.21  & 0.61  & 0.24  & 0.87  & 0.29  & 1.21  & 0.33  & 1.66 \\
    \bottomrule
    \end{tabular*}
    
\begin{mycaption}
The sensitivity ($\delta$ and $hd$) of the real-world networks with respect to systematic vertex removal is listed in the table above. We observe relatively small changes across all
social networks (the first three graphs listed above). In contrast, all web graphs are very sensitive to vertex removal induced by certain removal strategies.
For every web graph, the sensitivity values regarding $\delta$ and $hd$  (at $\theta = 0.30$)  for the most effective removal strategy are shown in bold.
Note, that with respect to $bc$ as removal strategy and $hd$ as comparison method, the sensitivity of the web graphs is larger than 6 while the sensitivity of the social networks does not exceed 0.52.
\end{mycaption}

  \label{tab:resultsEmpirical}%
\end{table*}

Since the majority of the $\delta$-values are monotonically increasing, we restrict the description of the results to $\theta = 0.3$ in most cases.

Considering social networks, we observe relatively small changes across all networks.
Except for $lp$, almost all removal strategies lead to similar behavior.
Since the underlying community detection algorithm may returns different communities in 
successive runs, the results referring to $lp$ should be treated with caution.
As a result, the removal order of the nodes changes. Especially in the cases of Google and Stanford, 
the ranking provided by $lp$ is unstable.

Excluding $lp$, which decreases in case of Hamsterster and is significantly lower than all other values in the remaining
cases, we observe the following mean (standard deviation) for $\delta$: Hamsterster 0.26 (0.060), Brightkite
0.45 (0.053), and Slashdot 0.27 (0.043).

Regarding web graphs, we note that some removal strategies substantially change the structure of the network. Specifically named,  
$dc_{out}$ (8.51), $bc$ (6.02), and $cc_{out}$ (3.83) in case of Google, $bc$ (241.81), $lp$ (36.89), $cc$ (35.60), $cc_{out}$ (22.23), and $pr$ 
(14.88) in case of NotreDame and $bc$ (14.18), $lp$ (8.49), and $ec$ (4.65) in case of Stanford. It should be noted that in instance of 
NotreDame the largest values of $cc$ ($cc_{out}$) already appear at $\theta = 0.25$ ($\theta = 0.15$). Moreover, the remaining removal 
strategies show higher $\delta$-values compared to social networks. However, we find that this behavior is diminished by symmetrization.\footnote{The $\delta$-values for the symmetrized web graphs are still at a higher level when compared to social networks but the effect is less noticeable. Taking Google (NotreDame) as an example, $\delta$ for $bc$ is lowered to 1.25 (5.17). Additional sensitivity values for symmetrized versions of the directed real-world networks can be found in Appendix~\ref{sec:appendix}.}

These findings are consistent with a  previous study by Boldi et al. \cite{Boldi2013a} who observed a difference in the behavior between social networks and web graphs.
They find that $bc$ and $lp$ are the most effective removal strategies with regard to web graphs.
Consistent with our observations, the mentioned study did not observe any significant changes with respect to the structure of social networks.

Additional to the harmonic diameter change, we also compare the neighborhood function by means of the relative average distance change
($\delta_{avg dist}$ ) and the percentage of reachable pairs ($\delta_{reachable}$). We notice the same behavior as in case of $\delta$. The 
values for the social networks increase moderately whereas all web graphs are significantly disturbed by some removal strategies. However,  in 
some cases the $\delta_{avg dist}$ (e.g. $bc$ for NotreDame and Stanford) increases first and decreases again with increasing $\theta$. 
Furthermore, the $\delta_{reachable}$ for undirected graphs only indicates how disconnected the graph is. These effects have also been 
observed in \cite{Boldi2013a} and therefore we only consider $\delta$ in the remainder of this study.

Using stochastic quantifiers is another way to compare $G$ and \gp.
The results for $hd$ are listed in Table~\ref{tab:resultsEmpirical}.
Comparing social networks and web graphs, fewer disturbances are shown for social networks whereas web graphs are considerably disturbed by some removal 
strategies. But these strategies are not necessarily the same as in instance of $\delta$:  $bc$, being the most efficient strategy to 
disturb NotreDame with respect to $\delta$, is only placed third in connection with $hd$ and $dc_{out}$ (Google) shows a $hd$ of 0.28 despite being ranked first regarding $\delta$.

All three stochastic quantifiers show similar results among each other. Although  $hd$ and $jsd$ are normalized and $kl$ is not, all measures behave similarly. Like Cabral et al. \cite{Cabral2014a}, we note that $hd$ is more sensitive to changes regarding the network structure compared to $jsd$, thus we focus our discussion on $hd$.

\begin{figure*}[htb]

\caption{Sensitivity of real-world networks with regard to centrality measures as comparison}

\label{fig:emp-cm-pr}

\includegraphics[width=\linewidth]{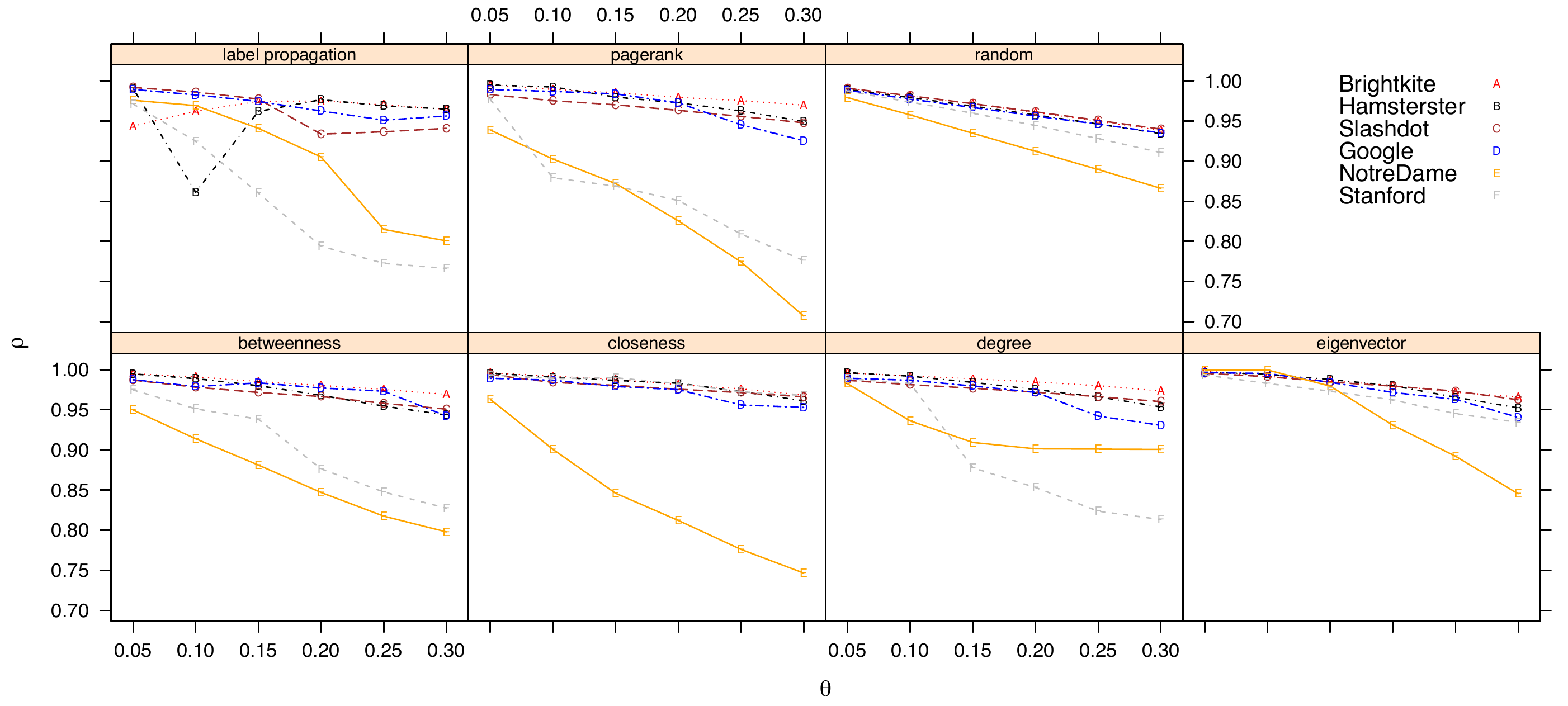}

\begin{mycaption}
The sensitivity ($pr$ as comparison method) of the real-world networks is illustrated in the figure above. Each removal strategy is represented by
one panel. We observe no clear-cut distinction between social networks (A, B, C) and web graphs (D, E, F) in any case.
\end{mycaption}

\end{figure*}

The results for $pr$ as comparison method are shown in Figure~\ref{fig:emp-cm-pr}. Among all centrality measures as removal strategy, we observe minor 
differences as far as the social networks and the google graph are concerned ($\rho \approx 0.95$ at $\theta = 0.30$). Although NotreDame and 
Stanford show increased sensitivity. The results of $dc$ as comparison method look similar but show less variation with a minimum $\rho$ of 0.80.
With regard to $ev$ as comparison method, Google shows lower values than the social networks with $bc$, $lp$ and $ev$ as removal strategy. Also 
in those cases there is no unambiguous discrimination between the two types of networks.
Overall, we find that no combination of removal strategy and centrality method is able to provide a clear-cut distinction between social 
networks and web graphs.

\section{Study of simulated networks}
In this section, we analyze the behavior of random graphs with respect to controlled modifications. We are especially interested in the following questions:

\textbf{Question 1:}
Do similarly generated random graphs show similar sensitivity to vertex removal?
Is the variance of the results of the simulations low enough to compare the values derived for different levels of $\theta$?

\textbf{Question 2:}
To what extend depends the sensitivity of random graphs on the parameterization? 
How do parameters i.e. the network size and other model-specific parameters influence the sensitivity?

\textbf{Question 3:}
What impact have the choice of removal strategy and comparison method on the sensitivity of random graphs? 
Is there a difference between the different removal strategies or measures of comparison with respect to the sensitivity?

\label{sec:generated}
\subsection{Experimental design and Data}
In contrast to the previous section, we compute the sensitivity values for a selection of simulated networks and
calculate the neighborhood function exactly in all cases.
We choose $\theta \in \{0.05, 0.1, 0.15, 0.2, 0.25, 0.3\}$ and $n \in \{2426, 15763\}$ which is the size of the Hamsterster, respectively Google, network.

To obtain comparable results, we choose overlapping parameters for $ER$ and $BA
$ and consider two different rewiring probabilities for both sizes of the $WS$ networks.

The graphs generated by $CF$ are based on the degree sequence of the Hamsterster, respectively Google, network. Therefore, we compare the generated graphs to their respective source graphs.
The specific parameters are shown in Table~\ref{tab:RandomStats}.
Since the label propagation community detection algorithm returns a single community for networks generated by $ER$ and $BA$ (as already 
mentioned in \cite{Raghavan2007a}), $lp$ is not considered in this section. Instead, we consider a random removal order as baseline model to make our 
results comparable to previous studies regarding the robustness of centrality measures.

\begin{table*}[t]
\centering
  \caption{Parameters and properties of networks generated by random graph models}
\begin{tabular}{lrrrr}
    \toprule
    model & n     & density & parameter & runs \\
    \midrule
    $ER$    & 2.426  & 0.0014, 0.0028, 0.0057, 0.0113, 0.0226 & $p$ = 0.0014, 0.0028, 0.0057, 0.0113, 0.0226 & 100 \\
          & 15.763 & 0.0003, 0.0006, 0.0012, 0.0024, 0.0048 & $p$ = 0.0003, 0.0006, 0.0012, 0.0024, 0.0048 & 10 \\
    $BA$    & 2.426  & 0.0049, 0.0058, 0.0066, 0.0074, 0.0082 & $l$ = 6, 7, 8, 9, 10 & 100 \\
          & 15.763 & 0.0008, 0.0009, 0.0010, 0.0011, 0.0013 & $l$ = 6, 7, 8, 9, 10 & 10 \\
    $WS$    & 2.426  & 0.0058, 0.0058 & $k$ = 7, $p_{rew}$ = 0.01, 0.16 & 100 \\
          & 15.763 & 0.0011, 0.0011 & $k$ = 9, $p_{rew}$ = 0.01, 0.16 & 10 \\
    $CF$    & 2.426  & 0.0055 & degree sequence of $Hamsterster$ & 100 \\
          & 15.763 & 0.0011 & degree sequence of $Google$ & 10 \\
    \bottomrule
    \end{tabular}
  \label{tab:RandomStats}

\end{table*}

\subsection{Results}
\begin{figure*}[htb]
\caption{Variance of similarly generated random graphs}
\label{fig:variance}
\includegraphics[width=\linewidth]{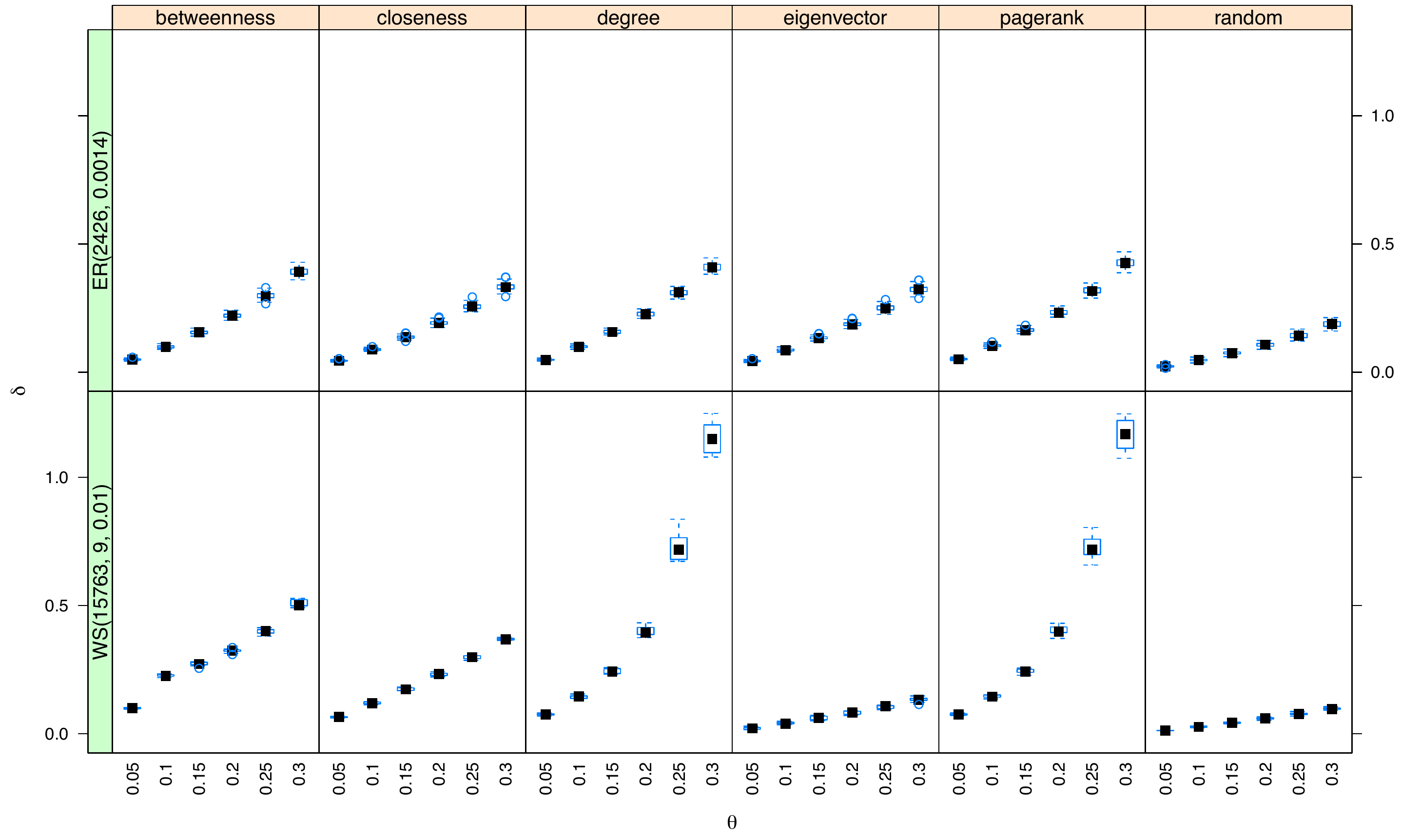}

\begin{mycaption}
The figure above illustrates examples for the variance of similarly generated random graphs in form of box-and-whisker plots ($\delta$ for $ER(2426, 0.0014)$ and $WS(15763, 9, 0.01)$).
Like in all other cases, we rarely see the range of the comparison methods overlap for two different modification levels.
\end{mycaption}

\end{figure*}

With respect to Question 1, we observe low standard deviations across all scenarios. Simulations based on $ER$ and $CF$ show continuously the 
lowest relative standard deviations. $WS$ and $BA$ show a higher variance but are still at an acceptable level. The variance for $ER(2426, 0.0014)$ 
and $WS(15763, 9, 0.01)$ are displayed as examples in Figure~\ref{fig:variance} in form of box-and-whisker plots. Other cases show similar behavior.
Due to the low variance, we rarely see the range of comparison methods overlap for two different $\theta$-values.
Since all measures are monotonically increasing, except some values concerning the centrality measures as comparison which we do mention 
separately, we focus our discussion on $\theta = 0.3$ in this section.

Our results regarding Question 2 and 3 are summarized in Table~\ref{tab:resultsGenerated} and described in detail in the next sections.

\begin{table*}[htb]
\caption{Summary of the results for generated graphs} 

\begin{tabular}{ll|p{7cm}p{7cm}}
\toprule

Graph &  & Q2: To what extend depends the sensitivity of random graphs on the parameterization? & Q3: What impact have the choice of removal strategy and comparison method on the sensitivity of random graphs? \\ 
\midrule

$ER$ & $N$ & The higher the $p$, the lower the sensitivity. Small graphs are more sensitive than large graphs. & There is only a difference between random and non-random removal. Both comparison methods behave similar. \\ 

 & $cm$ & Higher $p$ leads to lower sensitivity. Size has no influence for $dc$ and $pr$. For $bc$, $cc$ and $ec$ the larger graph is less sensitive. & Non-random removal strategies show similar behavior. The sensitivity differs between the comparison methods. $pr$, $dc$, $bc$, $cc$, and $dc$ become (in this sequence) more sensitive. \\ 
 \hline
$BA$ & $N$ & Similar to $ER$ regarding network size. A higher $l$ leads to lower sensitivity but size makes no difference for random vertex removal. & Similar to $ER$. \\ 

 & $cm$ & No difference regarding network size and $l$ for $pr$ and $dc$ as comparison methods and for random as removal strategy. In the remaining cases: sensitivity decreases with increasing network size and $l$. & Similar to $ER$. \\ 
 \hline
$WS$ & $N$ & Graphs with $p_{rew} = 0.01$ are more sensitive than graphs with $p_{rew} = 0.16$. Network size has a small influence on the sensitivity. & Graphs with $p_{rew} = 0.16$ show the same sensitivity for all removal strategies. For networks with $p_{rew} = 0.01$, random and $ec$ show low, $bc$ and $cc$ show medium, $pr$ and $dc$ show large sensitivity. Both sensitivity measures behave similarly. \\ 
 & $cm$ & Similar as for the neighborhood case. & Graphs with $p_{rew} = 0.16$ are similar to $ER$. For $p_{rew} = 0.01$, we hardly observe any patterns except for $dc$ and $pr$ as comparison method. They show similar behavior. \\ 

\end{tabular}
\newline

\begin{mycaption}
The table above shows a summary of the results for this section. As the variance is small in all cases, Question 1 is omitted for reasons of brevity. The results for graphs generated by $CF$ are not shown in the table above because we compare them to their respective source graph.
$N$ ($cm$)  denotes the comparison methods based on the neighborhood function (centrality measures).
\end{mycaption}
\label{tab:resultsGenerated}
\end{table*}

\subsubsection{Erd\H{o}s-R\'enyi model}
\begin{figure*}
\caption{Sensitivity of $ER(2426, p)$ regarding the neighborhood function}
\label{fig:er_nbhood}
\includegraphics[width=\linewidth]{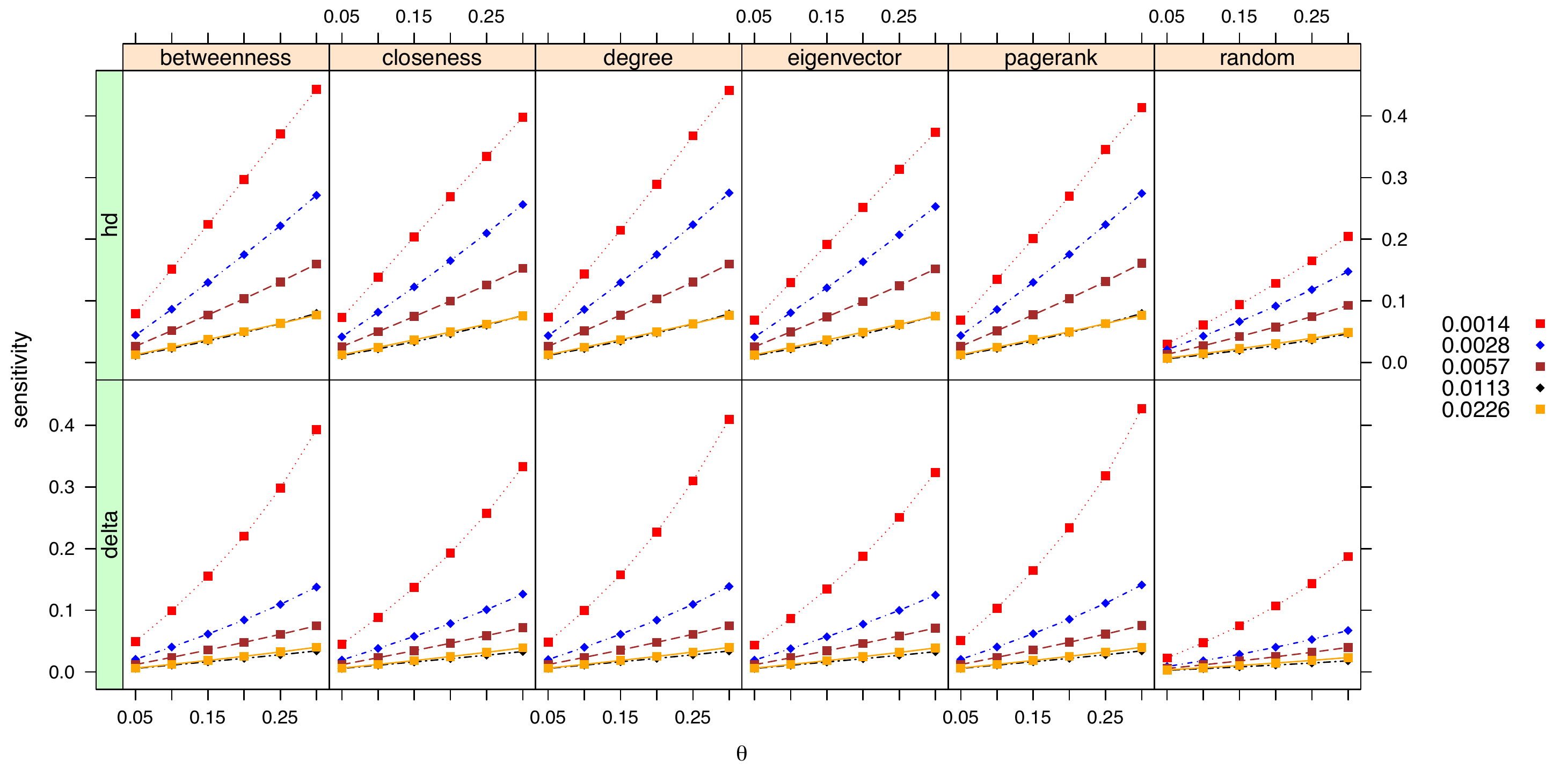}

\begin{mycaption}
The sensitivity of $ER$ with $n = 2426$ for different levels of $p$ is illustrated in the figure above. Different symbols are used for different levels of $p$. Both comparison methods, $\delta$ and $hd$,
show similar behavior at the respective level of $p$. Moreover, the sensitivity increases with decreasing $p$. With regard to the removal strategies, we observe that there is only a difference between random and non-random vertex removal. The sensitivity only differs slightly among non-random removal strategies.
\end{mycaption}

\end{figure*}

First, we take a look at the graphs generated by $ER$.
Based on the neighborhood function, the smaller graphs show higher values for $
\delta$ and $hd$ than the larger graph at the respective level of $p$. For example, the $ER(2426,0.0014)$ shows a $\delta$ of 0.36 whereas  
$ER(15763, 0.0012)$  shows a $\delta$ of 0.055. However, both sizes show the same behavior: the higher the $p$, the lower the $\delta$ or $hd
$. Comparing $hd$ with $\delta$, we notice that both measures behave in the same way (see Figure \ref{fig:er_nbhood}). Since this is the case for 
$BA$ and $WS$ as well, we subsequently focus on the behavior of $\delta$.

Regarding the removal strategies, we observe two different situations. Either we choose a random removal strategy and obtain a $\delta$ of 0.18 
for $ER(2426,0.0014)$, respectively 0.12 for $ER(15763, 0.0012)$, or we choose a removal strategy based on a centrality measure and obtain $\delta$ = 0.38 
\textpm 0.046 (0.23 \textpm 0.020).\footnote{These values represent the mean \textpm standard deviation of $\delta$ for all non-random removal strategies.}
In other words, there is a noticeable difference if the removal strategy is random or not but it barley makes a 
difference which non-random removal strategy is used. This effect appears for all combinations of $p$ and $\theta$ (Figure~\ref{fig:er_nbhood}).

Using the correlation of centrality measures as indicator for the structural change, we observe two of the already mentioned effects: with 
increasing $p$ the sensitivity decreases and all removal strategies based on centrality measures behave similarly whereas random removal has less 
impact on the structure. In contrast to $hd$ and $\delta$, the centrality measures differ among themselves. 
Unaffected by the network size and the removal strategy, $dc$ ($\rho = 0.89 \pm 0.0135$) and $pr$ ($\rho = 0.91 \pm 0.0128$) show the 
strongest correlation.

The remaining measures behave differently regarding the network size. Excluding the random removal, we observe $\rho$ = (0.85 \textpm 
0.0030, 0.77  \textpm 0.0135, 0.76 \textpm 0.0148) for $bc$, $cc$, $ec$ for $ER(15763, 0.0012)$, respectively $\rho$ = (0.78 \textpm 0.0206, 
0.60  \textpm 0.0437, 0.51  \textpm 0.0573) for $ER(2426,0.0014)$.

\subsubsection{Barabasi-Albert model}
Considering our results regarding graphs generated by BA, we observe similar findings. For a fixed $l$, graphs with $n = 2426$ 
show larger values for $\delta$ and $hd$ than graphs with $n = 15763$. The sensitivity decreases with increasing $l$;  $\delta$ and $hd$ show the same behavior with 
increasing $\theta$.
With regard to the sensitivity, there is little difference among all  non-random removal strategies. The random vertex removal consistently shows the lowest 
sensitivity and does not differ regarding the size of the graph.

Using $pr$ and $dc$ as comparison method, we observe similar behavior for all combinations of $l$ and $n$ in instance of all removal 
strategies. This behavior also is observed for $bc$, $cc$, and $ec$ in case of random vertex removal. For non-random removal strategies, 
these comparison methods differ. They show lower correlation for the larger graph and the correlation increases with increasing $l$.

Compared to $BA$-graphs, $ER$-graphs are less sensitive when centrality measures are used as comparison method. This is also true for $
\delta$ and $hd$ with one exception: If vertices are removed randomly, $BA$-graphs are less sensitive.

\subsubsection{Watts-Strogatz model}
When analyzing $WS$-graphs, we observe that graphs with $p_{rew} = 0.01$
are more sensitive than graphs with $p_{rew} = 0.16$. In both cases, there is little difference in sensitivity with respect to the network size.
For graphs with $p_{rew} = 0.16$, we note the same level of sensitivity for all removal strategies, including random vertex removal. Graphs with the lower 
rewiring probability behave differently. These observations are similar when considering both comparison methods, based on the neighborhood function and  on centrality measures.
Considering $\delta$  for graphs with $p_{rew} = 0.01$ ($hd$ behaves similar, again), vertex removal based on a random order and $ec$ has little impact ($\delta \approx 0.11$). $bc$ 
and $cc$ have 
medium impact ($\delta \approx 0.50$) and $dc$ and $pr$ have the largest impact ($\delta \approx 1.20$).

As far as centrality measures are used as comparison method, we notice two different situations. For $p_{rew} = 0.16$, the $WS$-graphs behave like 
$ER$-graphs in terms of sensitivity. The sensitivity is similar for all non-random removal strategies, except for $ec$ as comparison method.
For $p_{rew} = 0.01$, we hardly observe any patterns except for $dc$ and $pr$, which show similar behavior. The network size has negligible  influence on the sensitivity.

\subsubsection{Configuration model}
Since the different sized graphs generated by $CF$ are based on different degree sequences, we do not compare them to each other. 
Rather we investigate the similarity between the generate graph and the respective source graph.

When the neighborhood function is used as comparison meth\-od, we find that generated graphs show similar sensitivity for all non-random removal strategies ($CF(Hamsterster)$: $\delta \approx 0.16$, $CF(Google)$: $\delta \approx 0.38$). Random vertex removal leads to lower sensitivity
($CF(Hamsterster)$: $\delta \approx 0.07$, $CF(Google)$: $\delta \approx 0.08$). The $hd$-values for all removal strategies and the $\delta$-values for $ec$ and random removal for $CF(Hamsterster)$ are essentially equivalent to those of the respective source graph. 
Except for random vertex removal, $CF(Google)$ and its respective source graph do not show any similarities.

The sensitivity with regard to comparison by means of centrality measures only differs between random and non-random vertex removal.
Considering both 
generated graphs, the behavior is similar compared to their respective source graph if $dc$ (and $cc$, $bc$ in case of  $CF(Hamsterster)$) is used as comparison method.

\section{Conclusions}
In this paper, we analyze the sensitivity of real-world networks and random graphs with respect to systematic vertex removal. We consider a 
variety of removal strategies and comparison methods.

When using the neighborhood function based comparison methods, web graphs show high sensitivity. In contrast, 
social networks show low sensitivity. This finding is consistent with previous observations made by Boldi et al. \cite{Boldi2013a}. However, 
no comparison method based on a centrality measure provides a clear-cut distinction between social networks and web graphs.

We examine graphs generated by four different random graph models. We observe that the 
smaller graphs exhibit higher sensitivity than the larger graphs.
Furthermore, the comparison methods based on the neighborhood function show similar behavior regarding the sensitivity.
However, centrality based methods do not.
Our experiments show, that there is a difference between a random removal order and removal strategies based on centrality measures. However, in 
the majority of the cases, it does make little difference which non-random removal strategy we choose.

In this paper, we focused on the systematic removal of vertices. Future research may investigate the sensitivity with respect to
systematic insertion of vertices and nodes as well as the behavior of directed graphs.
Another step towards a better understanding of the sensitivity of web graphs might be the usage of exponential random graph models, in order to
simulate networks that share various structural properties with the respective source graph.

\bibliographystyle{abbrv}
\bibliography{StructuralAspectsNetworks} 
\clearpage
\newpage

\appendix
\section{Additional data}
\label{sec:appendix}

\begin{table}[htb]
  \centering
  \caption{Sensitivity ($\delta$) for symmetrized versions of the directed real-world networks}
    \begin{tabular}{llrrrrrr}
    \toprule
      $G$    &   \diagbox{$R$}{$\theta$}    & 0.05  & 0.10  & 0.15  & 0.20  & 0.25  & 0.30 \\
    \midrule
    Slashdot & $bc$ & 0.07  & 0.10  & 0.13  & 0.18  & 0.22  & 0.28 \\
          & $cc$ & 0.05  & 0.10  & 0.13  & 0.16  & 0.18  & 0.24 \\
          & $dc$ & 0.08  & 0.11  & 0.14  & 0.16  & 0.20  & 0.26 \\
          & $ec$ & 0.04  & 0.09  & 0.09  & 0.15  & 0.19  & 0.23 \\
          & $lp$ & 0.04  & 0.08  & 0.12  & 0.06  & 0.02  & 0.01 \\
          & $pr$ & 0.06  & 0.12  & 0.16  & 0.18  & 0.22  & 0.29 \\
          \midrule
          
    Google & $bc$ & 0.25  & 0.37  & 0.81  & 0.90  & 1.09  & 1.25 \\
          & $cc$ & 0.25  & 0.37  & 0.40  & 0.49  & 0.70  & 0.97 \\
          & $dc$ & 0.25  & 0.37  & 0.40  & 0.49  & 0.60  & 1.03 \\
          & $ec$ & 0.25  & 0.37  & 0.38  & 0.49  & 0.55  & 0.65 \\
          & $lp$ & 0.25  & 0.46  & 0.69  & 0.86  & 0.94  & 1.09 \\
          & $pr$ & 0.25  & 0.48  & 0.62  & 0.90  & 1.09  & 1.30 \\
          \midrule
          
    NotreDame & $bc$ & 0.43  & 0.86  & 1.36  & 2.11  & 3.27  & 5.17 \\
          & $cc$ & 0.34  & 0.70  & 1.63  & 4.21  & 8.78  & 17.43 \\
          & $dc$ & 0.05  & 0.32  & 0.48  & 0.80  & 1.18  & 1.64 \\
          & $ec$ & 0.03  & 0.04  & 0.03  & 0.01  & 0.00  & 0.25 \\
          & $lp$ & 0.24  & 0.58  & 1.11  & 2.41  & 4.70  & 7.84 \\
          & $pr$ & 0.29  & 0.54  & 0.73  & 1.09  & 1.71  & 2.70 \\
          \midrule
          
    Stanford & $bc$ & 0.29  & 0.69  & 1.16  & 1.50  & 1.83  & 2.55 \\
          & $cc$ & 0.25  & 0.29  & 0.31  & 0.38  & 0.56  & 0.69 \\
          & $dc$ & 0.24  & 0.22  & 0.53  & 0.53  & 0.80  & 0.93 \\
          & $ec$ & 0.24  & 0.33  & 0.43  & 0.30  & 0.42  & 0.52 \\
          & $lp$ & 0.28  & 0.65  & 1.48  & 2.35  & 2.98  & 3.81 \\
          & $pr$ & 0.26  & 0.48  & 0.65  & 0.75  & 0.85  & 1.01 \\
    \bottomrule
    \end{tabular}
  \label{tab:addlabel}
\end{table}

\end{document}